\documentclass[%
 reprint,
 amsmath,amssymb,
 aps,
 prl,
 longbibliography,
 lengthcheck,%
]{revtex4-1}

\usepackage{graphicx}% Include figure files
\usepackage{dcolumn}% Align table columns on decimal point
\usepackage{bm}% bold math
\usepackage{hyperref}% add hypertext capabilities
\begin{document}

\preprint{APS/123-QED}

\title{Hubbard fermions band splitting at the strong intersite Coulomb interaction}% Force line breaks with \\

\author{Valery V. Val'kov}
 \email{vvv@iph.krasn.ru}
\author{Maxim M. Korovushkin}%
 \email{maxim.korovushkin@gmail.com}
\affiliation{%
 L.V. Kirensky Institute of Physics, 660036 Krasnoyarsk, Russia\\}%

\date{\today}% It is always \today, today,

\begin{abstract}
The effect of the strong intersite Coulomb correlations on the
formation of the electron structure of the $t-V$--model has been
studied. A qualitatively new result has been obtained which
consists in the occurrence of a split-off band of the Fermi
states. The spectral intensity of this band increases with the
enhancement of a doping level and is determined by the mean-square
fluctuation of the occupation numbers. This leads to the
qualitative change in the structure of the electron density of
states.
\begin{description}
\item[PACS numbers]
71.10.Fd, 71.18.+y, 71.27.+a, 71.28.+d, 71.70.-d
\end{description}
\end{abstract}

\pacs{Valid PACS appear here}% PACS, the Physics and Astronomy
\maketitle

%\tableofcontents

The key point of the theory of the strongly correlated electron
systems is the statement on a principal role of the one-site
Coulomb repulsion of two electrons with the opposite spin
projections, e.g., the Hubbard correlations~\cite{H63}, in the
formation of the ground state and the elementary excitation
spectrum~\cite{Anderson}. One of the brightest manifestations of
the Hubbard correlations is splitting of the initial band of the
energy spectrum into two Hubbard subbands when one-site energy of
the Coulomb interaction $U$ exceeds the bandwidth $W$.

As was noted previously~\cite{Rogdai1,Rogdai2}, when hole
concentration in a system described by the Hubbard model is small
and value $U$ is large the intersite Coulomb interaction starts
playing an important role due to its relatively weak screening. In
this case at distances close to interatomic the order of magnitude
of characteristic value $V$ of the Coulomb interaction between
electrons can be comparable with that of value $U$. Manifestation
of the intersite correlations in the physical properties of
systems has been considered in many works (see, for instance,
\cite{Rogdai1,Rogdai2,Falicov,Khomskii,Matsukawa,Littlewood,Fedro,Li,Larsen,Onishi,Neudert,Barreteau,Miyake}).
However, until recent time no attention has been paid to the fact
that the strong intersite correlations (SIC) can cause splitting
energy bands of the Fermi states into subbands~\cite{VVMK}.

Qualitatively, the physical origin of this phenomenon is similar
to that of the occurrence of the Hubbard subbands due to the
strong one-site correlations and is related to the fact that when
the intersite Coulomb interactions are taken into account the
energy of the electron located on site $f$  becomes dependent of
the valence configuration of its nearest neighborhood. For example
in the case of a lower Hubbard subband, if there is one electron
on each of the $z$ nearest neighboring ions, the "setting" energy
of an electron is determined as a sum of one-site energy
$\varepsilon$ and energy $zV$ of the interaction. If a hole
appears in the nearest neighborhood, the "setting" energy is
$\varepsilon+(z-1)V$, i.e., its value is less than the previous
one by $V$.

These simple arguments show that when configuration neighborhood
deviates from nominal one, one should expect the occurrence of the
states with the energies different by $V$ in the Fermi excitation
spectrum. For the strongly correlated systems with a relatively
narrow energy band of the Fermi states, the situation becomes
possible when hopping parameter $t$ is commensurable with or less
than $V$. Under these conditions splitting the initial band of the
Fermi states is expected. Obviously, the more probable the
deviation of the electron configuration of the neighborhood from
the nominal one the higher is the spectral intensity of the
split-off band. A quantitative measure of this deviation is a
mean-square fluctuation of the occupation numbers. For this reason
the split-off band was named the band of the fluctuation states
(BFS)~\cite{VVMK}. For undoped Mott-Hubbard insulators the
electron configuration of the neighborhood corresponds to the
nominal one. Upon doping the deviation occurs and in the energy
spectrum of the Fermi states the BFS appears whose spectral
intensity increases with doping level. The aim of this Letter is
to prove the presented qualitative interpretation.

In order to demonstrate the effect as brightly as possible we will
consider the Hubbard model in the regime of the strong electron
correlations ($U=\infty$) at electron concentrations $n\ll1$. In
this case the electron properties of the model are determined by
the lower Hubbard subband. Considering the arguments presented
in~\cite{Rogdai1,Rogdai2} we will take into account the Coulomb
interaction between the electrons located on the neighboring
sites. The obtained system of the Hubbard fermions in the atomic
representation will be described by the Hamiltonian of the
$t-V$--model:
\begin{equation}
\label{Ham} \hat{H}=\sum\limits_{f\sigma}\varepsilon_0
X_f^{\sigma\sigma}+\sum\limits_{fm\sigma}t_{fm}X_f^{\sigma0}X_m^{0\sigma}+\frac
{V}{2}\sum\limits_{f\delta}\hat n_f\hat n_{f+\delta}.
\end{equation}
Here the first term reflects an ensemble of noninteracting
electrons in the Wannier representation. The occurrence a fermion
with spin projection $\sigma$ at site $f$ increases the energy of
the system by value $\varepsilon_0$, $X^{pq}_f=|f,p\rangle\langle
f,q|$ are the Hubbard operators~\cite{H65,SGOVV} describing the
transition from the one-site state $|q\rangle$ to the state
$|p\rangle$. The second term corresponds to the kinetic energy of
the Hubbard fermions where matrix element $t_{fm}$ determines the
intensity of electron hopping from site $f$ to site $m$. The last
term of the Hamiltonian takes into account the Coulomb interaction
between the electrons located on neighboring sites $f$ and
$f+\delta$ with intensity $V$. The operator of the number of
electrons on site $f$ is $\hat n_f=\displaystyle\sum_{\sigma}
X_f^{\sigma\sigma}$.

Below we limit our consideration to the case when the number of
holes $h=(1/N)\sum_f\langle X_f^{00}\rangle$ in the system is
small; i. e., the condition $h=1-n\ll1$ is satisfied. In this
regime it is reasonable to extract in explicit form the mean-field
effects caused by the intersite interactions. Using the condition
of completeness of the diagonal $X$-operators in the reduced
Hilbert space
$X^{00}_f+X^{\uparrow\uparrow}_f+X^{\downarrow\downarrow}_f=1$ we
express Hamiltonian (\ref{Ham}) as
\begin{eqnarray}\label{Cor_VHam}
\hat H &=&-E_0+
\sum\limits_{f\sigma}\bigl(\varepsilon-4Vh\bigr)X_f^{\sigma\sigma}
+\sum\limits_{fm\sigma}t_{fm}X_f^{\sigma0}X_m^{0\sigma}
\nonumber\\
&&+\frac12\sum_{f\delta}V\bigl(X^{00}_f-h\bigr)\bigl(X^{00}_{f+\delta}-h\bigr),
\end{eqnarray}
where $E_0=2NV(1-h)^2$ in the mean-field approximation determines
the energy of the Coulomb interaction of the system containing $h$
holes per site. At $h=0$ value $E_0$ equals to the exact value of
the energy of the ground state of the system (with disregard for
$\varepsilon_0$), since in this case hoppings make no
contribution. The renormalized value of a one-electron level
$\varepsilon=\varepsilon_0+4V$ is determined by the fact that when
there is one electron on each neighboring site the excitation
energy increases by $4V$. The shift $\varepsilon-4hV$ is related
to the decrease in the Coulomb repulsion energy when the average
number of holes in the system is nonzero. Note that such
mean-field renormalizations of the one-site energies of electrons
were used previously, for example, in the Falicov--Kimball model
during investigation of the transitions with the change in
valence. Extraction of the obvious mean-field effects is needed
for representation of the intersite interaction in the form
containing the correlation effects only. One can see that the last
term of the Hamiltonian (\ref{Cor_VHam}) will contribute only at
the presence of noticeable fluctuations of the occupation numbers,
i. e., when the SICs are relatively large.

The method used below for the description of the strong intersite
interactions (SII) consists in generalization of the Hubbard
idea~\cite{H63} for consideration of the intersite
interactions~\cite{VVMK}. It follows from the exact equation of
motion for operator $X_f^{0\sigma}$
\begin{eqnarray}\label{eq1}
i\frac{d}{dt}X^{0\sigma}_f&&=(\varepsilon-4Vh)X^{0\sigma}_f-V
\sum_{\delta}X_f^{0\sigma}(X^{00}_{f+\delta}-h)\nonumber\\
&&+\sum_m t_{fm}\bigl((X_f^{00}+X_f^{\sigma\sigma})X_m^{0\sigma}+
X_f^{\bar{\sigma}\sigma}X_m^{0\bar{\sigma}}\bigr),
\end{eqnarray}
that at large $V$ one should correctly consider the contributions
related to the second term in the right part of equation
(\ref{eq1}). One of the ways of solving such a problem is to
extend the basis set of operators by means of inclusion of the
operators
\begin{equation}\label{phi}
\hat{\Phi}_f^{\sigma}=\sum_{\delta}X_f^{0\sigma}(X^{00}_{f+\delta}-h),
\end{equation}
in which uncoupling is forbidden. This operator describes the
correlated process of annihilation of an electron on site $f$,
since the result of the action of operator $X_f^{0\sigma}$ depends
on the nearest configuration neighborhood of site $f$. Recall for
comparison that in the fundamental Hubbard's work~\cite{H63} the
basis was expanded by means of addition of a set of one-site
operators $a_{f\sigma}\hat{n}_{f\bar{\sigma}}$.

Inclusion of new basis operators requires the equations of motion:
\begin{eqnarray}\label{eq2}
&&i\frac{d}{dt}\hat \Phi^{\sigma}_f=
(\varepsilon-2Vh-V)\hat\Phi^{\sigma}_f-4h(1-h)VX_f^{0\sigma}
\nonumber\\
&&+\sum_{m\delta} t_{fm}
\Bigl((X_f^{00}+X_f^{\sigma\sigma})X_{m}^{0\sigma}+
X_f^{\bar{\sigma}\sigma}X_{m}^{0\bar{\sigma}}\Bigr)
(X_{f+\delta}^{00}-h)\nonumber\\
&&+\sum_{m\delta\sigma'}t_{f+\delta,m}X^{0\sigma}_f
(X^{\sigma'0}_mX^{0\sigma'}_{f+\delta}-X^{\sigma'0}_{f+\delta}X^{0\sigma'}_m)
\nonumber\\
&&-V\sum_{\delta(\delta\neq\delta_1)}
X_f^{0\sigma}(X_{f+\delta}^{00}-h)(X_{f+\delta_1}^{00}-h).
\end{eqnarray}
One can see that among the operators containing large parameter
$V$ a three-site operator has occurred, which reflects the
correlation effects related to the presence of two holes in the
first coordination sphere of site $f$. If hole concentration in
the system is low, then the contribution of this term can be
ignored. Calculation of the spectrum with regard of this
three-center operator confirmed validity this approximation.

After the basis set of operators has been specified, the set of
equations of motion is closed using the Zwanzig-Mori projection
technique~\cite{Zwanzig,Mori}. In the main approximation one can
neglect the kinetic correlators occurring after projection and
limit the consideration to spatially homogeneous solutions. Then,
for the Fourier transforms of the Green functions we obtain the
closed system of two equations:
\begin{eqnarray}\label{system}
(\omega-\varepsilon_{\bf k})\langle\langle X_{{\bf
k}\sigma}|X_{{\bf k}\sigma}^\dagger\rangle\rangle &=&
\frac{1+h}{2}-\gamma_{\bf k}\langle\langle \Phi_{{\bf
k}\sigma}|X_{{\bf
k}\sigma}^\dagger\rangle\rangle,\\
(\omega-\xi_{\bf k})\langle\langle \Phi_{{\bf k}\sigma}|X_{{\bf
k}\sigma}^\dagger\rangle\rangle &=& -4h(1-h)\gamma_{\bf
k}\langle\langle X_{{\bf k}\sigma}|X_{{\bf
k}\sigma}^\dagger\rangle\rangle.\nonumber
\end{eqnarray}
Here we made the following notation:
\begin{eqnarray}
&&\varepsilon_{{\bf
k}}=(\varepsilon-4Vh)+\biggl(\frac{1+h}{2}\biggr)t_{\bf k},
\quad\gamma_{\bf k}=V-\frac{t_{1{\bf k}}}{8},\nonumber\\
&&\xi_{\bf k}=(\varepsilon-4Vh)-V(1-2h)+\biggl(\frac{2+h}{8}\biggr)t_{1{\bf k}},\nonumber\\
&&t_{\bf k}=t_{1{\bf k}}+4t'cos\,k_xcos\,k_y+
2t''(cos\,2k_x+cos\,2k_y),\nonumber\\
&&t_{1{\bf k}}=2t(cos\,k_x+cos\,k_y).
\end{eqnarray}
The use of the spectral theorem allows obtaining the equation
which determines the dependence of the chemical potential on
doping:
\begin{eqnarray}\label{selfcons}
\frac{4h}{1+h}&=&\frac{1}{N} \sum_{\bf k\rm}
\left(1+\frac{\varepsilon_{\bf k}-\xi_{\bf k}}{2\nu_{\bf
k}}\right) \left(1-\emph{f}\;(E^+_{\bf k})\right),
\end{eqnarray}
where $f(x)=(\exp(\frac{x-\mu}{T})+1)^{-1}$ is the Fermi-Dirac
function, $\mu$ is the chemical potential of the system, and the
two-band Fermi excitation spectrum is determined as
\begin{eqnarray}\label{spectrum}
&&E^{\mp}_{\bf k}=\frac{\varepsilon_{\bf k}+\xi_{\bf k}}{2}\mp
\nu_{\bf k},\\
&&\nu_{\bf k}= \sqrt{\biggl(\frac{\varepsilon_{\bf k}-\xi_{\bf
k}}{2}\biggr)^2+ 4h(1-h)\gamma^2_{\bf k}}.\nonumber
\end{eqnarray}

\begin{figure}[htbp]
\begin{center}
\includegraphics[width=0.41\textwidth]{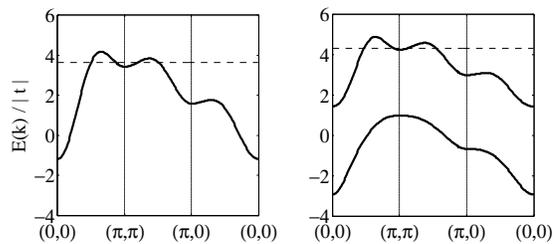}
\caption{Band picture of the $t-V$--model without account for the
SICs (left) and with account for the SICs (right) for the set of
parameters $t'=-0.1,\, t''=-0.5,\, V=2.5$ (in terms of $|t|$) and
holes concentration $h=0.15$. Dashed lines show the chemical
potential.} \label{bands}
\end{center}
\end{figure}

Fig.\ref{bands} (on the right) demonstrates a band picture of the
energy spectrum of the $t-V$--model obtained by solving equations
(\ref{system}). The values of hopping parameters were chosen so
that evolution of the Fermi contour upon hole doping would
qualitatively correspond to that observed experimentally. The left
part of the figure shows the energy spectrum of the same model
calculated with disregard of the SICs. Note that here the
mean-field contribution of the SIIs is taken into account by the
above-mentioned renormalizations. Comparison of the two presented
spectra shows that the correct account for the SICs yields the
qualitative difference, specifically, the occurrence of an
additional band (BFS) in the band structure of the $t-V$--model.

It is seen from (\ref{spectrum}) that the resulting energy
spectrum forms by hybridization of states from an ordinary Hubbard
band with energies $\varepsilon_{\bf k}$ and the states induced by
the fluctuations of configuration neighborhood with energies
$\xi_{\bf k}$. Note that unlike $\varepsilon_{\bf k}$ the
dependence of function $\xi_{\bf k}$ on the quasimomentum is
determined only by the integral of hopping between the nearest
neighbors $t_{1{\bf k}}$. At $V\gg|t_{\bf k}|$ the values of this
function form an energy band located lower by $V$. The intensity
of hybridization is determined by the value proportional to the
one-site mean-square fluctuation of the occupation numbers
\begin{equation}
\overline{(\Delta
n)^2}=\langle(X_f^{00}-h)(X_f^{00}-h)\rangle=h(1-h).
\end{equation}
As a result the spectral intensity of the BFS in the regime
$V\gg|t_{\bf k}|$ acquires the simple form:
\begin{equation}
A^-({\bf k},\omega)\simeq \frac{4\left(1+h\right)\overline{(\Delta
n)^2}}{1+16\overline{(\Delta n)^2}+\sqrt{1+16\overline{(\Delta
n)^2}}}\delta(\omega-E^{-}_{\bf k}).
\end{equation}
It follows from this formula that with an increase in $h$ the
spectral weight of the BFS rapidly grows and at $h=0.2$ the
relative contribution of the BFS reaches 30\%.
\begin{figure}[htbp]
\begin{center}
\includegraphics[width=0.31\textwidth]{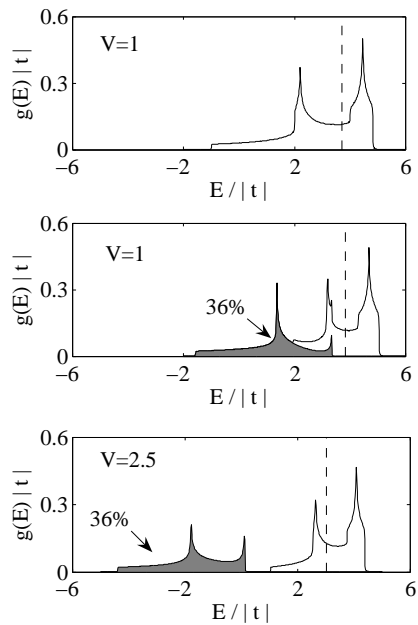}
\caption{Density of states of the $t-V$--model calculated for
$h=0.25$ without account for the SICs $V=1$ (top), with account
for the SICs at $V=1$ (middle) an at $V=2.5$ (bottom). Here
$t'=-0.1,\, t''=-0.5$ (all in terms of $|t|$). Dashed lines show
the chemical potential.} \label{DOS}
\end{center}
\end{figure}
\begin{figure*}[htbp]
\begin{center}
\includegraphics[width=0.53\textwidth]{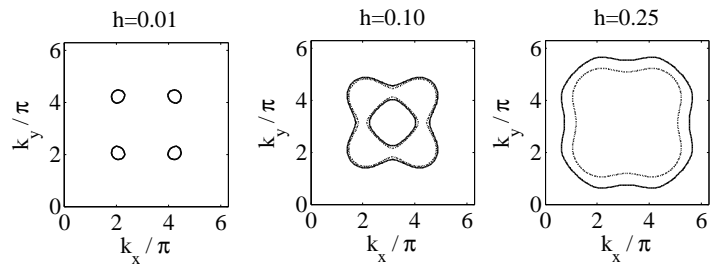}
\caption{Fermi surfaces calculated for the set of parameters
$t'=-0.1,\, t''=-0.5,\, V=2.5$ ( in terns of $|t|$) and different
$h$ without account for the SICs (dashed lines) and with account
for the SICs (solid lines).} \label{FS}
\end{center}
\end{figure*}

The occurrence of the BFS qualitatively changes the electron
density of states of the Hubbard fermions. The upper graph in
Fig.\ref{DOS} shows the density of states of the $t-V$--model with
disregard of the SICs. It is seen that ignoring these correlations
leads to trivial displacement of the band position.

The energy dependence of the sum electron density of states of the
$t-V$--model for the same set of parameters, but with regard of
the SICs, is shown by a bold solid line in the middle graph of
Fig.\ref{DOS}. Comparison with the upper graph evidences that the
account for the SICs yields noticeable qualitative changes in the
density of states: the latter acquires a three-peak structure
instead of the two-peak one. The occurrence of the additional peak
corresponds to the formation of the BFS. For clarity, BFS density
$g^-(E)$ is reflected in the graph by the line which bounds the
shaded area. The BFS fraction is 36\% of the whole number of
states of the system, whereas the fraction of the ground band with
density of states $g^+(E)$ is 64\%. Summation of densities
$g^+(E)$ and $g^-(E)$ gives total density $g(E)$ with the three
peaks.

More substantial changes in the energy structure of the
$t-V$--model take place at high values of the intersite Coulomb
interaction. It is seen from the lower graph in Fig.\ref{DOS}
which presents the same dependences as in the two upper graphs but
calculated for $V=2.5$ (the rest parameters are not changed). In
this case, the SICs lead to splitting the BFS off and the
formation of a gap in the energy spectrum. The graphs presented in
Fig.\ref{DOS} are calculated at $h=0.25$. In the case of $h=0$ the
SIC contribution becomes zero and the density of states of the
system will be such as shown in the upper graph in Fig.\ref{DOS}.
This suggests the presence of a qualitatively new effect related
to the account for the SIC: upon doping not only the chemical
potential shifts but the density of states rearranges.

The growth of the BFS spectral intensity with an increase in
doping level leads to renormalization of the dependence of the
chemical potential on hole concentration in the system. As a
result in the area of optimal doping $h\simeq0.2$ the square
bounded by the Fermi contour noticeably grows. The Fermi contour
calculated at $h=0.25$ with disregard of the SICs is shown by
dotted lines in Fig.\ref{FS}. If the correlations are taken into
account then the Fermi contour noticeably grows and acquires the
form shown by solid lines. In this case the change in the square
bounded by the Fermi contour reaches $16\%$ (right graph). This
increase is important for description of the Lifshitz quantum
phase transitions occurring upon doping~\cite{SGO} and for
interpretation of the experimental data on measuring magnetic
oscillations in the de Haas--van Alphen effect. Recently, such
measurements have been performed by many researchers due to
substantial improvement of quality of materials and novel
techniques with the use of the strong magnetic fields.

Note in summary that the hybridization character of the energy
spectrum with regard of the SICs is related to the fact that
electron hoppings between the nearest neighbors lead to the
transitions between the one-site energy levels which differ by
$V$. Therefore, the intensity of such hybridization processes is
proportional to both hopping parameter $t$ and the above-mentioned
mean-square fluctuation of the occupation numbers. Note also that
the considered modification of the energy structure due to the
SICs is general and not limited merely by the $t-V$--model. One
should expect that the effects discovered in this study will be
especially important for the systems with variable valence.

This study was supported by the program "Quantum physics of
condensed matter" of the Presidium of the Russian Academy of
Sciences (RAS); the Russian Foundation for Basic Research (project
No. 07-02-00226); the Siberian Division of RAS (Interdisciplinary
Integration project No. 53). One of authors (M.K.) would like to
acknowledge the support of the Dynasty Foundation.

\end{document}